\documentclass{jfm}
\usepackage{graphicx}
\usepackage{epstopdf, epsfig}

\usepackage{enumitem}

\usepackage{wrapfig}
\usepackage{units}

\usepackage{orcidlink}
\hypersetup{
  colorlinks   = true,  %Colours links instead of ugly boxes
  urlcolor     = blue,  %Colour for external hyperlinks
  linkcolor    = blue,  %Colour of internal links
  citecolor   = blue    %Colour of citations
}

\newcommand{\beq}{ \begin{equation} }
\newcommand{\eeq}{ \end{equation} }
\newcommand{\beqs}{ \begin{eqnarray} }
\newcommand{\eeqs}{ \end{eqnarray} }

\newcommand{\vect}[1]{\ensuremath{%
        {\mbox{\mathversion{bold}\ensuremath{#1}}}%
        }}

\usepackage{amsmath}
\usepackage[ISO]{diffcoeff}  
\usepackage{units}

\usepackage{xcolor}

\setlist[itemize]{leftmargin=5.5mm}

\usepackage{tcolorbox}
\newtcolorbox{mybox}[3][]
{
  colframe = #2!25,
  colback  = #2!10,
  #1,
}

\usepackage{blindtext}

\shorttitle{Modeling vorticity effects in respiratory transport}
\shortauthor{S. Basu, L.P. Chamorro, M. Yeasin and M.A. Stremler}

\title{Modeling the effect of vorticity on\\inhaled transport in the upper airway}
\author{S. Basu\aff{1}
  \corresp{\email{Saikat.Basu@sdstate.edu}},
  L. P. Chamorro\aff{2},
 M. Yeasin\aff{1}
  \and M. A. Stremler\aff{3}
 }

\affiliation{\aff{1} Mechanical Engineering, South Dakota State University, Brookings, South Dakota 57007
\aff{2}Mechanical Science and Engineering, University of Illinois, Urbana, Illinois 61801
\aff{3}Mechanical Engineering, Virginia Tech, Blacksburg, Virginia 24061
}

\begin{document}

\maketitle

%\noindent\rule{\textwidth}{0.5pt}\\

\begin{abstract}
Localized vortices can have significant influence on transport of inhaled particles through the upper respiratory tract.  These vortices have complex three-dimensional structure with details dependent on the anatomical geometry.  Using a highly simplified model, we demonstrate that changes in transport characteristics with geometric distortion can be estimated by accounting merely for the net strength and location of the vorticity in a two-dimensional projection.  Test cases consider 30\,L/min inhaled airflow containing suspended spherical water droplets from 1\,$\mu$m to 30\,$\mu$m in diameter through (1)~a healthy upper respiratory tract and (2)~a distorted variation mimicking a glottic tumor.  The reduced-order model approximates the system by a two-dimensional potential flow with embedded point vortices having features derived from Large Eddy Simulations of inhaled airflow through  anatomically realistic, tomography-based, three-dimensional tracts. The effects of vorticity and particle size on changes in particle transport are shown to be consistent between the reduced-order model and the full-scale simulations.  
\end{abstract}

%\begin{keywords}
%Respiratory transport; Vortex dynamics; Reduced-order modeling; Biofluid mechanics; Upper respiratory tract
%\end{keywords}

%%%%%%%%%%%%%%%%%%%%%%%%%%%%%%%%%%%%%%%%%%%%%%%%%%%%%%%%%%%%%%%%%%%%%%%%%%%%%%%%%%%%%%%%%%%%%%%%%%%%%%%%%%%%%%%%%%%%%%%%%%%%%%%%%%%%%%%%%%%%%%%%%%%%%%%%%%%%%%%%%%%%%%%%%%%%%%%%%%%%%%%

\section{Introduction}\label{intro}
\begin{figure}
\begin{center}
\includegraphics[width=\textwidth]{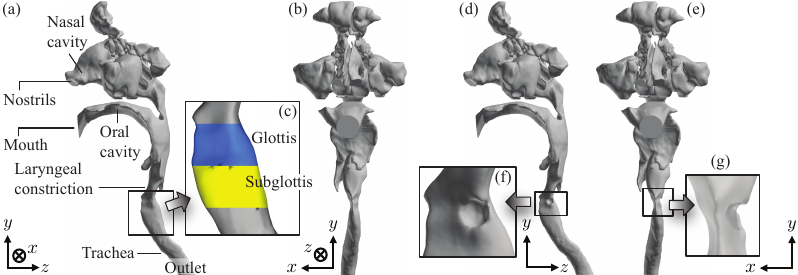}
\caption{(a) Sagittal view of an anatomically accurate 3D healthy human upper respiratory tract, reconstructed from CT scans. (b) Coronal view of the same healthy tract. The glottis and subglottis are highlighted in panel (c) in blue and yellow, respectively. Panels (d) and (e) demonstrate the sagittal and coronal views of the tumor-embedded tract, with zoomed-in visuals of the tumor shown in panels (f) and (g). Rendering in (g) is partially transparent for visibility.}
\label{f1}
\end{center}
\end{figure}

Vortices formed within the upper respiratory tract play a crucial role in particle transport dynamics during breathing. Inhaled air interacts with the tortuous  airway geometry (e.g., figure~\ref{f1}), particularly around bends and constrictions, leading to shear layer separation and emergence of coherent vortex patches; see e.g., \citet{kleinstreuer2010arfm,yuk2022jrs,moriarty1999jfm}. Such vortices can significantly influence the local dispersion and deposition of inhaled particles. 
For instance, simulations of inhaled airflow almost always reveal strong vortex structures just downwind of the vocal folds in the throat as the airway widens following a laryngeal constriction \citep{perkins2018ohns}. Depending on their size and inertia, particles may be trapped within or slowed by the low-pressure vortex cores or redirected toward the airway walls, affecting their overall trajectories and deposition sites.

Modeling and characterizing the vortex-particle interactions are pivotal for systematically describing particle behavior in respiratory health and disease, e.g., influencing intra-airway therapeutic delivery and the transmission of inhaled pollutants, including allergens and pathogens. While case-by-case numerical simulations and experimental tests in subject-specific or idealized test geometries are the established approaches (see e.g., \citet{inthavong2019jb,basu2021scirep,akash2023fdd}), these methods are challenging and time-consuming. A critical question remains: How feasible is to use a simple vortex dynamics model to replicate the strong influence of vorticity on particle dynamics, particularly in the throat region?

% Figure 1

To address this question, we have developed a reduced-order mathematical model (ROM) that consists of a two-dimensional (2D) potential flow with embedded point vortices, slip boundary conditions, and particle tracking using a simplified representation of the Maxey-Riley equations.  Details of the model are presented in \S\ref{theory}.  The features and parameters of the ROM are informed by full-scale, three-dimensional (3D) Large Eddy Simulations (LES) of inhaled transport in anatomical reconstructions of the upper respiratory tract, as described in \S\ref{sim}.   We focus on the flow-induced particle deposition trends immediately downstream of the laryngeal constriction in the glottis and subglottis regions (see figure~\ref{f1}a,c). 

As discussed in \S\ref{results}, the relationship between particle size and changes in transverse particle motion in the theoretical model is found to be consistent with the relationship between particle size and deposition in the glottic and subglottic regions from the high-fidelity computational model. 
These results suggest that it is the overall magnitude and extent of vorticity in the upper respiratory tract, not the specific details of vortex structure and orientation, that primarily controls particle deposition in the throat.  The implications of this work are summarized in \S\ref{discussion}.
 
%The efficacy of the 
%particle size dependence for glottic and subglottic deposition from the theoretical model and the numerical tests are found remarkably consistent, highlighting the potential to expand the novel analytical framework for wider parametric analysis of transmission trends within the throat and, more broadly, through the entire upper respiratory tract.

%%%%%%%%%%%%%%%%%%%%%%%%%%%%%%%%%%%%%%%%%%%%%%%%%%%%%%%%%%%%%%%%%%%%%%%%%%%%%%%%%%%%%%%%%%%%%%%%%%%%%%%%%%%%%%%%%%%%%%%%%%%%%%%%%%%%%%%%%%%%%%%%%%%%%%%%%%%%%%%%%%%%%%%%%%%%%%%%%%%%%%%

\section{Full-scale computational analysis}\label{sim}

% Figure 2
\begin{figure}
\begin{center}
\includegraphics[width=\textwidth]{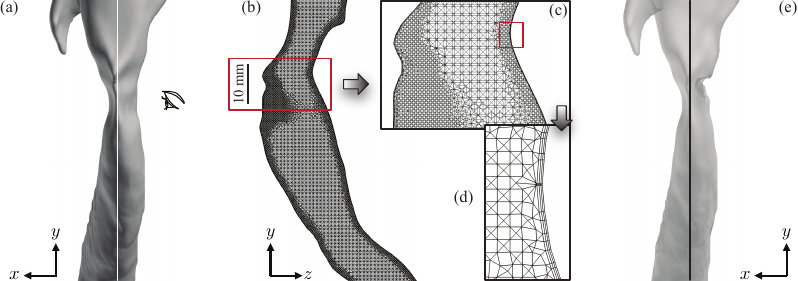}
\caption{(a) Cropped coronal view of the healthy tract. (b-d) Progressive zoomed-in views of the 3D mesh cross-section. Red box in (b) shows the accurate length scale.  (e) Cropped coronal view in the tumor-embedded tract. White and black lines in (a,e) mark the cut-away in (b). }
\label{f2}
\end{center}
\end{figure}

High-fidelity computational simulations were conducted in anatomically realistic airways using the LES scheme with a dynamic subgrid-scale kinetic energy transport model to simulate inhaled air flux at 30 L/min. The simulated airflow was combined with Lagrangian discrete phase particle transport analysis to determine intra-airway deposition and penetration patterns. 
%These computational results guided definition of the ROM parameters in \S\ref{theory}.

\subsection{In silico test geometries and airspace discretization}

The computational simulations considered a healthy upper respiratory tract (see figure~\ref{f1}a-c) built from high-resolution computed tomography (CT) imaging and its digital twin (see figure~\ref{f1}d-g) with an anatomical distortion in the form of a spherical nodule (hereafter referred to as a ``tumor'') near the vortex emergence site. Tract geometries were reconstructed from medical-grade CT slices collected at coronal depth increments of approximately 0.4\,mm using radiodensity thresholding between -1024 and -300 Hounsfield units \citep{basu2018ijnmbe,borojeni2017ijnmbe}. To examine the effect of anatomical perturbations on inhaled transport, the tumor-embedded domain was digitally prepared by placing a 6-mm diameter spherical distortion at the glottis of the healthy tract.  This tumor mimics the location and size of a clinically realistic laryngeal granuloma. 
The resulting geometries were spatially discretized into over six million unstructured, graded tetrahedral elements following established mesh refinement protocols (see, e.g., \citet{frank2016jampdd}. To resolve particle dynamics near the walls, three layers of pentahedral cells, each with a height of 0.033\,mm and an aspect ratio of 1.1, were extruded along the cavity surfaces. See figure~\ref{f2} for multi-scale visualizations of the resulting mesh.

% Figure 3
\begin{figure}
\begin{center}
\includegraphics[width=\textwidth]{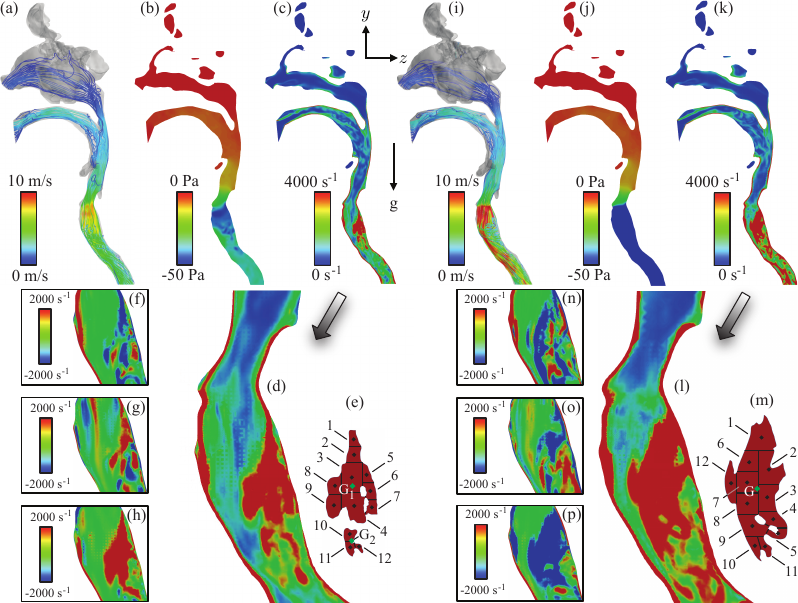}
\caption{Numerical simulation data for 30 L/min inhaled airflow, showing (a,i)~representative velocity streamlines, (b,j)~pressure fields, and (c,k)~vorticity magnitude contours. Panels (a-h) correspond to the healthy tract and (i-p) correspond to the tumor-embedded tract. Panels (d,l) call out the vorticity-dominated regions from (c,k). Insets (e,m) show vortex patch segregation. Black circles indicate geometric centers of each segregated patch (numbered and separated by dark lines) and the green circles mark the geometric centers (G$_\textrm{k}$) of the entire patch areas. (f-h) and (n-p) show the $x$, $y$, $z$-vorticity components. g marks the gravity direction in the simulations.
}
\label{f3}
\end{center}
\end{figure}

\subsection{Computational simulation of inhalation air flow}

Assuming incompressible and isothermal flow, the filtered  continuity and momentum equations are
\begin{equation}\label{e1}
    \frac{\partial}{\partial x_i}\left(\rho \overline{u}_i\right) = 0,
\end{equation}
\begin{equation}\label{e2}
    \frac{\partial \overline{u}_i}{\partial t} + \frac{\partial}{\partial x_j}\left(\overline{u}_i\overline{u}_j\right) = - \frac{1}{\rho_f}\frac{\partial \overline{p}}{\partial x_i} + \frac{\partial}{\partial x_j}\left(\nu\frac{\partial \overline{u}_i}{\partial x_j}\right) - \frac{\partial \tau_{ij}}{\partial x_j},
\end{equation}
where $\overline{u}_i$ is the resolved velocity, $\overline{p}$ is the resolved pressure, $\nu$ is the kinematic viscosity of air, $\rho_f$ is the air density, and $\tau_{ij}$ is the subgrid-scale stress tensor that is modeled as
\begin{equation}\label{e3}
    \tau_{ij} - \frac{1}{3}\tau_{kk}\delta_{ij} = -2\, \nu_{s}\,\overline{S}_{ij}.
\end{equation}
Here, $\delta{ij}$ represents Kronecker delta, $\nu_{s}$ is the subgrid-scale eddy viscosity, and $\overline{S}_{ij}$, as the rate-of-strain tensor for the resolved scale, is defined as
\begin{equation}\label{e4}
    \overline{S}_{ij} = \frac{1}{2}\left(\frac{\partial \overline{u}_i}{\partial x_j} + \frac{\partial \overline{u}_j}{\partial x_i}\right).
\end{equation}
The isotropic component of the subgrid-scale stresses, i.e., $\tau_{kk}$, is not modeled but added to the filtered static pressure term. To account for the transport of subgrid-scale turbulent kinetic energy in the LES scheme, which can be prominent in complex flow domains such as the human respiratory tract, we applied the dynamic subgrid-scale kinetic energy model \citep{kim1997ketm} 
described by $\nu_{s} = C_k \, \tilde{\Delta}\, k^{1/2}_{s}$. Here, $C_k$ (and $C_{\varepsilon}$ in \eqref{e7}) are model constants determined dynamically. $\tilde{\Delta}$ is the filter characteristic length, computed as the cube root of the grid cell volume, and $k_s$ is the subgrid-scale kinetic energy obtained by solving the filtered transport equation
\begin{equation}\label{e7}
    \frac{\partial k_{s}}{\partial t} + \frac{\partial}{\partial x_j}\left(\overline{u}_j k_{s}\right) = \frac{\partial}{\partial x_j}\left(\nu_{s}\frac{\partial k_{s}}{\partial x_j}\right) + \frac{\partial \overline{u}_i}{\partial x_j}\left(2\nu_{s}\overline{S}_{ij} - \frac{2}{3}k_{s}\delta_{ij}\right) - \frac{C_{\varepsilon}\, k^{3/2}_{s}}{\tilde{\Delta}}.
\end{equation}
The simulated 30 L/min airflow required a driving mean pressure difference of 35.16\,Pa in the healthy tract and 73.43\,Pa in the tumor-embedded tract between the inlet (at the left and right nostrils and mouth) and the outlet (at the tracheal base). Time step was set at $2 \times 10^{-4}$ s \citep{ghahramani2017cf}. The no-slip boundary condition was imposed at the tissue and cartilage surfaces enclosing the geometries. The density and kinematic viscosity of the tracked fluid (air) were set at 1.204\,kg/m$^3$ and $1.825 \times 10^{-5}$\,m$^2$/s. The simulations used a segregated solver with pressure-velocity coupling and second-order upwind spatial discretization. Figure~\ref{f3} outlines the flow simulation data.
%Simulated values for the material properties were kept consistent with those subsequently used in the ROM.

\subsection{Computational simulation of inhaled particle transport}
One-way coupling was used between the solved airflow and the tracked particles, in which the particles are carried by the streamlines but do not impact the surrounding flow. 
The underlying flow field was assumed (quasi-)steady while computing the particle transport. A Lagrangian-based inert discrete phase model was used to numerically integrate the particle transport equation,
\begin{equation}\label{e8}
   \frac{d v_{pi}}{dt} = \frac{18\mu \,C_D Re_p \left(u_i - v_{pi}\right) }{24 \rho_p d^2} + g_i\left(1 - \frac{\rho_f}{\rho_p}\right) + B_i,
\end{equation}
where $v_{pi}$ is the particle velocity, $u_i$ is the airflow velocity, $\mu$ is the molecular viscosity of the background fluid (air), $C_D$ is the coefficient of drag \citep{morsi1972jfm}, $Re_p$ is the particle-based Reynolds number, $\rho_p=1$\,g/mL is the material density of the tracked particles (assuming water droplets), $g_i$ is the gravitational acceleration component, and $B_i$ represents additional body forces per unit particle mass, such as the Saffman lift force exerted by a flow shear field on small particles transverse to the ambient flow direction. The simulations tracked monodisperse particle sets with diameters $d \in [1,\ 30]$ $\mu$m in increments of $1\,\mu$m, with $N_p = 8002$ and $5972$ particles of each size tracked in the healthy tract and the tumor-embedded tract, respectively. $N_p$ is a function of the test geometry and represents the net number of surface mesh elements on the reconstructed mouth and nostrils. The geometric centers of these elements served as the initial positions for the simulated particles. Tracking of a particle was terminated once it reached the mesh element layer adjacent to the cavity walls, and the corresponding coordinates were recorded as its deposition location. 

\begin{figure}
\begin{center}
\includegraphics[width=\textwidth]{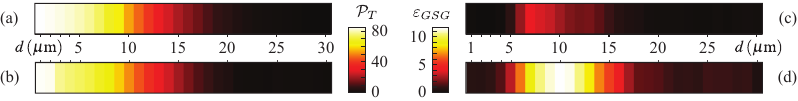}
\caption{Computationally simulated transport of particles initially entering the nostrils and mouth. (a,b) Probability, $\mathcal{P}_{T}$ (in \%),  of initial particles reaching the glottic airspace in the (a)~healthy and (b)~tumor-embedded tracts. (c,d)~Percentage of initial particles deposited in the glottis or subglottis regions, $\varepsilon_{GSG}$, for the (c)~healthy and (d)~tumor-embedded tracts.     }
\label{f4}
\end{center}
\end{figure}

Figure~\ref{f4} summarizes the particle transport data.  Panels (a,b) quantify the  probability $\mathcal{P}_T$ that a particle entering through the nostrils or mouth will enter the glottic region, i.e., without undergoing prior deposition on the wall of the airway. The $\mathcal{P}_T$ trend is visibly similar across the two geometries, with a standard deviation of only 1.94\% between the $\mathcal{P}_T$ values for these two flows across each simulated particle size.  For particles larger than $d \gtrsim \unit[18]{\mu m}$, less than 10\% of all tracked particles reached the airspace at the glottis. 

In contrast, particle deposition in the target region was significantly different between the two flow geometries.  Figure~\ref{f4}c,d shows how flow geometry and particle size affect the deposition efficiency, $\varepsilon_{GSG} = (\mathcal{N}_{GSG}/N_p) \times 100$, 
where $\mathcal{N}_{GSG}$ is the number of particles deposited at the glottis and subglottis in each particle simulation. 
%Taking a cut-off  level of 2\%, the preferred size range for glottic and subglottic deposition is $\unit[6]{\mu m} \lesssim d \lesssim \unit[13]{\mu m}$ in the healthy tract and $\unit[5]{\mu m} \lesssim d \lesssim \unit[17]{\mu m}$ in the tumor-embedded tract. 
Glottic and subglottic deposition is greater than 2\% for $\unit[6]{\mu m} \lesssim d \lesssim \unit[13]{\mu m}$ in the healthy tract and $\unit[5]{\mu m} \lesssim d \lesssim \unit[17]{\mu m}$ in the tumor-embedded tract.

% Figure 4

\subsection{Extraction of flow features}
\begin{figure}
\begin{center}
\includegraphics[width=\textwidth]{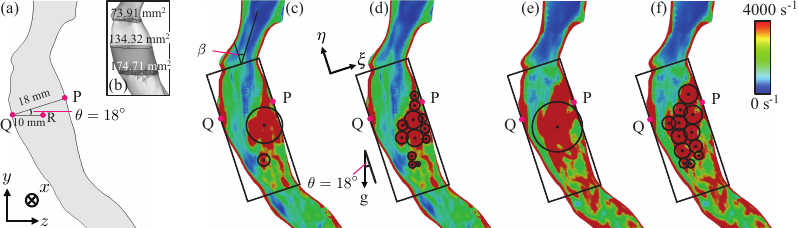}
\caption{(a,b)~Geometric features of the cut-away cross-section and airspace. 
(c-f)~ROM representations (solid lines and dots) of the vorticity magnitude field (color maps) are shown for the (c)~low and (d)~high resolution models of the healthy tract and for the (e)~low and (f)~high resolution models of the tumor-embedded tract. Solid rectangles show the ROM domain. Solid dots show point vortex locations $\zeta_i$; open circles show the corresponding areas $A_i$. Points P, Q, R provide reference locations for spatial comparison. }
\label{f5}
\end{center}
\end{figure}

As shown in figure~\ref{f3}a--k, the simulated flow fields are similar between the healthy tract and the tumor-embedded tract prior to the glottic airspace. Correspondingly, figure~\ref{f4}a,b shows that the probability of particles entering the glottic region, $\mathcal{P}_T$, is essentially independent of anatomical geometry (for the cases considered here).  
Strong 3D vortices appear in both geometries in the glottic and subglottic regions, but the vorticity generated in the tumor-embedded geometry is distinctly stronger.  Figure~\ref{f4}c,d shows that there is a distinct difference in particle deposition in these regions between the healthy and tumor-embedded tracts.  
Our assumption was that the primary influence of the vortical structures on particle transport is to drive particles away from the paths of passive particles and thus toward the throat wall, with the effect being dependent on particle size but largely independent of vortex orientation.  The net influence of vorticity in this flow was thus approximated simply by taking the magnitude of vorticity in the 2D cross-section indicated in figure~\ref{f2}a,e, with a focus on the boxed region in figure~\ref{f5}.   The vorticity-dominated region in each geometry was segregated into 12 smaller patches (see figure~\ref{f3}e,m), with the mean vorticity magnitude and patch area determined for each (see table~\ref{t1}). The properties of the full coherent patches were also determined for an alternative `low-resolution' model.

 % Figure 5 

\begin{table}
\begin{center}
\begin{tabular}{rrrlrl}
& Patch  & \multicolumn{2}{l}{Healthy tract} & \multicolumn{2}{l}{Tumor-embedded tract} \\
&     ID            & $A_i$ (mm$^2$)~~       & $\omega_i$ (s$^{-1}$)~~~~~~       & $A_i$ (mm$^2$)~~       & $\omega_i$ (s$^{-1}$)        \\[3pt]
high-resolution & 1~~          & 6.10~~                & 4000~~~~~~                        & 31.52~~               & 4000              \\
model & 2~~          & 9.24~~                & 4000~~~~~~                        & 29.36~~               & 4000              \\
& 3~~         & 27.08~~               & 4000~~~~~~                        & 21.73~~               & 4000              \\
& 4~~          & 24.63~~                & 4000~~~~~~                       & 13.93~~               & 4000              \\
& 5~~          & 6.51~~                & 4000~~~~~~                        & 6.65~~                & 4000              \\
& 6~~          & 7.58~~                & 4000~~~~~~                        & 16.69~~               & 4000              \\
& 7~~          & 5.99~~                & 4000~~~~~~                        & 20.81~~               & 4000              \\
& 8~~          & 9.88~~                & 4000~~~~~~                        & 20.47~~               & 4000              \\
& 9~~          & 13.15~~               & 4000~~~~~~                        & 21.04~~               & 4000              \\
& 10~~          & 5.83~~                & 4000~~~~~~                        & 8.60~~                & 4000              \\
& 11~~          & 4.42~~                & 4000~~~~~~                        & 6.37~~                & 4000              \\
& 12~~          & 2.05~~                & 3500~~~~~~                        & 15.73~~               & 4000              %\\[3pt]
%$\sum A_i$ & 122.451           &               & 212.902              &               
\\[3pt]
low-resolution & G1/G & 110.15~~ & 4000 & 212.90~~  & 4000 \\
model & G2~~ & 12.30~~ & 3900 & ---~~~~ & ~~---
\end{tabular}
\caption{Vortex patch parameters; see figure~\ref{f3} for the patch IDs.  
%{\color{red} [ADD $\zeta_i$ values??]} 
}\label{t1}
\end{center}
\end{table}

%%%%%%%%%%%%%%%%%%%%%%%%%%%%%%%%%%%%%%%%%%%%%%%%%%%%%%%%%%%%%%%%%%%%%%%%%%%%%%%%%%%%%%%%%%%%%%%%%%%%%%%%%%%%%%%%%%%%%%%%%%%%%%%%%%%%%%%%%%%%%%%%%%%%%%%%%%%%%%%%%%%%%%%%%%%%%%%%%%%%%%%

\section{A simplified mathematical model}\label{theory}

The reduced-order model (ROM) consisted of a 2D potential flow through a straight channel with point vortices placed to simulate the vorticity distribution in the computational flow. The influence of this induced flow on particle transport was modeled using a simplified version of the Maxey-Riley equation \citep{maxey1983pf}.

\subsection{Reduced-order flow model}
The ROM represents the throat section within the rectangle in figure~\ref{f5}c-f, with flow parameters guided by the full-scale computational simulations in \S\ref{sim}. Based on measurements from the CT-derived reconstructions (e.g., figure~\ref{f5}a), the model channel width is $W = \unit[18]{mm}$ and the channel is oriented at $\theta = 18^\circ$ from vertical. Air at standard conditions was assumed to flow steadily at a rate of $Q = 30$ L/min. Assuming a representative cross-sectional glottis area of $A = \unit[130]{mm^2}$ (see figure \ref{f5}b), the background channel speed was taken to be $U = \unit[3.8]{m/s}$. The transported particles were assumed to be water droplets with diameters ranging from $d = 1$ to $\unit[30]{\mu m}$.

The flow was modeled in the dimensionless complex $\zeta$ plane, with $\zeta = \xi + \mathrm{i}\,\eta$ and $\mathrm{i}\,\eta$ aligned with the channel axis (see figure~\ref{f5}). Channel walls were located at $\xi = 0$ and $\xi = 1$, and the channel length was $L/W = 2.5$. The vortical structures were represented by fixed point vortices, with locations and circulations determined from the computational simulations.  
We considered both a `high resolution' and a `low resolution' model to examine the influence of vortex details on particle transport.  In the high-resolution model, each of the 12 vortex patches in figure~\ref{f3} was mimicked by a single point vortex placed at the patch's geometric center $\zeta_i$ (see figure~\ref{f3}e for the healthy tract and figure~\ref{f3}m for the tumor-embedded tract), with dimensionless circulations given by $\Gamma_i = \omega_i A_i / UW$, where $\omega_i$ is the mean vorticity in the patch area $A_i$; see table~\ref{t1}.   For the low-resolution model, the full coherent vortices patches were represented by point vortices (two vortices for the healthy tract and one vortex for the tumor-embedded tract).  

% Figure 6
\begin{figure}
\begin{center}
\includegraphics[width=\textwidth]{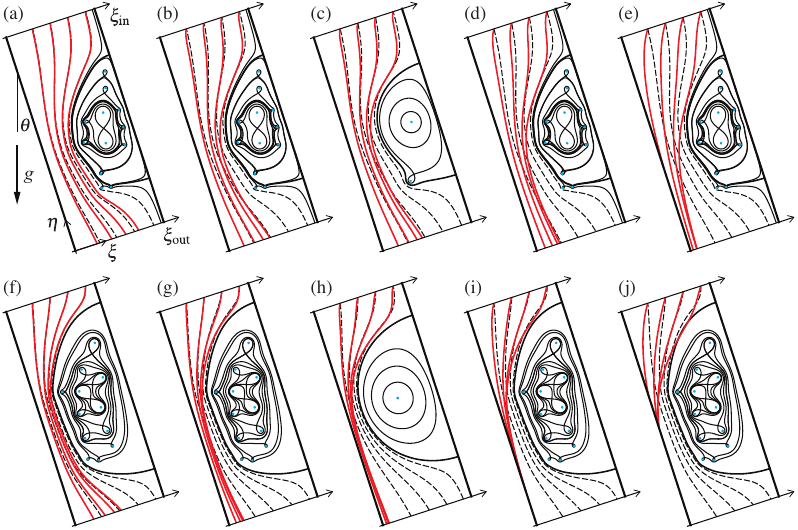}
\caption{Streamlines (black lines) and particle pathlines (red lines) from the ROM with slip wall boundaries (heavy black lines). Flow is from top to bottom.  Sample particles have $\xi_{in} \in \{0.3, 0.5, 0.7, 0.9\}$ and $\beta=35^{\circ}$.  
Point vortex locations are marked by solid blue circles. 
Streamlines are shown passing through the particles’ initial positions (dashed lines), connecting the stagnation points at $\xi=1$ that define the primary ``vortex bubble'' (medium solid lines), and passing through the stagnation points in the flow (thin solid lines) for the (a-e)~healthy tract, (f-j)~tumor-embedded tract, and (c,h)~low-resolution vortex model, with all other panels the high-resolution model.  Trajectories are shown for particles with diameters $d$ as follows: (a, f) $\unit[5]{\mu m}$, (b,c,g,h)~$\unit[10]{\mu m}$, (d,i)~$\unit[15]{\mu m}$, and (e,j)~$\unit[25]{\mu m}$.
}
\label{f6}
\end{center}
\end{figure}

Straight streamlines along the channel walls were established by including periodic images of the point vortices in the $\xi$ direction. The resulting dimensionless complex potential for this flow, with a dimensionless background speed of unity, is given by \citep{friedmann1928fortschreitende}
\begin{equation}\label{e:potential}
    F(\zeta) = \phi(\xi, \eta) + \mathrm{i}\,\psi(\xi, \eta) = \mathrm{i}\, \zeta +  \sum_{i=1}^{N} \frac{\Gamma_i}{2\pi\mathrm{i}} \log \left\{ \frac{\sin\left[ \pi (\zeta-\zeta_i)/2 \right] }{\sin\left[ \pi (\zeta + \zeta_i^*)/2 \right]} \right\} , 
\end{equation}
with $\psi(\xi,\eta)$ being the real-valued flow streamfunction; the asterisk denotes complex conjugation.  Representative streamlines are shown in figure~\ref{f6}.

\subsection{Reduced-order  particle transport model}
Inhaled particle motion was modeled using a simplified version of the Maxey-Riley equation \citep{maxey1983pf}. In 2D, this equation can be written in vector form as \citep{babiano2000prl}
\begin{equation}\label{e:maxeyriley}
 \frac{d\vect{w}}{dt} = - \left[ \vect{J} + \frac{2\, {\text{St}}^{-1}}{3(\sigma+1)} \vect{I} \right] \cdot \vect{w} + \frac{\sigma}{\sigma+1} \left( {\text{Fr}}^{-2} \vect{g} - \frac{\text{D}\vect{u}}{\text{D}t} \right),
\end{equation}
where $\vect{u}(\vect{\xi}, t)$ is the local fluid velocity in vector coordinates $\vect{\xi} = (\xi, \eta)$, $\vect{w} = \vect{v} - \vect{u}$ is the relative velocity of a particle with velocity $\vect{v}$, $\vect{J}$ is the 2D Jacobian matrix, \vect{I} is the identity matrix, and $\vect{g}$ is the direction of gravity. The nondimensional parameters are
\begin{equation}
  \text{St} \equiv \frac{d^2 U}{18 \nu L}, \quad \text{Fr} \equiv \frac{U}{\sqrt{g L}}, \quad \text{and} \quad \sigma \equiv \frac{2}{3}\left(\frac{\rho_p}{\rho_f} -1 \right),
\end{equation}
for particles and fluid (air) with material densities $\rho_p$ and $\rho_f$. 
This model neglects the Faxen correction terms and the Basset-Boussinesq history force. Although viscous effects are neglected in equation \eqref{e:potential}, particle drag is included in equation \eqref{e:maxeyriley}.

Particles were taken to enter the ROM domain with speed $U$ and an initial angle $\beta = $ 25$^\circ$, 35$^\circ$, and 45$^\circ$ with respect to the $\eta$~axis (see figure~\ref{f5}c), which assumes that particles entering this region of the glottis were guided by the upwind physiological shape.  Representative particle trajectories are shown in figure~\ref{f6}.

%%%%%%%%%%%%%%%%%%%%%%%%%%%%%%%%%%%%%%%%%%%%%%%%%%%%%%%%%%%%%%%%%%%%%%%%%%%%%%%%%%%%%%%%%%%%%%%%%%%%%%%%%%%%%%%%%%%%%%%%%%%%%%%%%%%%%%%%%%%%%%%%%%%%%%%%%%%%%%%%%%%%%%%%%%%%%%%%%%%%%%%

\section{Assessment}\label{results}

Although our ROM approach in \S\ref{theory} was guided by the computational flow data in \S\ref{sim}, these two systems are significantly different, and a direct comparison of, say, particle deposition is not instructive.  To compare these two models, and thereby develop a deeper understanding of how vorticity affects particle transport in the upper airway, we examine changes in transport characteristics between the healthy and tumor-embedded tracts.  

\begin{figure}
\begin{center}
\includegraphics[width=\textwidth]{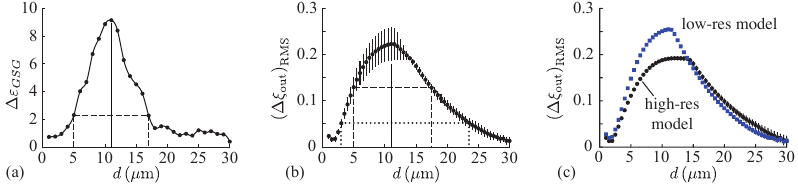}
\caption{(a) Deposition difference, $\Delta \varepsilon_{GSG}$, from the computational model  as a function of inhaled particle sizes. (b, c) Root mean square (RMS) of the exit difference $\Delta \xi_{\text{out}}$ in the ROM for (b) both low- and high-resolution vortex models and $\beta = \{25^{\circ}, 35^{\circ}, 45^{\circ}\}$.  Panel (c) separates the data for low- and high-resolution vortex models. Errors bars show one standard deviation.  
}
\label{f7}
\end{center}
\end{figure}

For the computational results, we consider the difference in deposition efficiency, $\Delta \varepsilon_{GSG} =  ( \varepsilon_{GSG} )_{\text{tumor}} -  ( \varepsilon_{GSG} )_{\text{healthy}}$, for the data presented in figure~\ref{f4}c,d.  The resulting variation as a function of particle size is shown in figure~\ref{f7}a.  
%For $d \lesssim \unit[5]{\mu m}$, particles are not prone to be deposited on the throat wall in either tract, with $\mathcal{P}_T$ high and $\varepsilon_{GSG}$ low for both. For $d \gtrsim \unit[17]{\mu m}$, $\mathcal{P}_T$  is low {\color{red} [QUANTIFY?] } for each tract, indicating that very few particles make it to the glottis in either case.  
The tumor-embedded geometry, and the associated increase in vorticity, causes the greatest change in particle deposition for $d = \unit[11]{\mu m}$.  Using $0.25 \Delta \varepsilon_{GSG} (\unit[11]{\mu m})$ as a threshold, the tumor-embedded tract has a significant influence on particle deposition for $\unit[5]{\mu m} \lesssim d \lesssim \unit[17]{\mu m}$.

In the ROM, the net effect of the vortices is to shift particles  in the negative $\xi$ direction. We tracked particles with $\unit[1]{\mu m} \le d \le \unit[30]{\mu m}$ (in increments of $\unit[0.5]{\mu m}$) for $N_R = 99$ initial conditions spread uniformly along $\xi_{\text{in}}$, with  $\xi_{\text{out}}$ recorded for each. 
The vortex-induced deviation in particle trajectories was quantified by
\begin{equation}
	(\Delta \xi_{\text{out}})_{\text{RMS}} = \sqrt{ \frac{1}{N_R} \sum_{j=1}^{N_R} \left[ \xi_{\text{out}} \!\left( \xi_{\text{in}, j} ; d, \beta \right)_{\text{tumor}} - \xi_{\text{out}} \!\left( \xi_{\text{in}, j} ; d, \beta \right)_{\text{healthy}} \right]^2
	}.
\end{equation}	
Results are shown in figure~\ref{f7}b,c, indicating that the RMS deviation is essentially independent of $\beta$.  While the details of the vortex model affect deviations for $\unit[5]{\mu m} \lesssim d \le \unit[14]{\mu m}$, this influence is primarily in magnitude rather than the overall trend.  

The greatest change in particle trajectories occurs for $d = \unit[11]{\mu m}$, corresponding exactly with the computational results.   Taking $0.25 (\Delta \xi_{\text{out}})_{\text{RMS}} (\unit[11]{\mu m})$ as a threshold shows a wider range of particle sizes, $\unit[3]{\mu m} \lesssim d \lesssim \unit[23.5]{\mu m}$, being affected by the increase in vorticity relative to the computational model.  A higher threshold gives particles with $\unit[5]{\mu m} \le d \le \unit[17.5]{\mu m}$ exhibiting significant change in transport, correlating well with the high-fidelity computational model.  

%Since the particles enter the domain at an angle with respect to the flow, their inertia causes a leftward deviation even without vortices in the flow. 
%The details of the particle motion vary subtly based on whether we use the high-resolution model, versus the low-resolution model; see the related comparison for 10-$\mu$m particle trajectories in figure~\ref{f6}(b, c) in the healthy tract and in figure~\ref{f6}(g, h) in the tumor-implanted tract.
%We have quantified the averaged leftward motion of the ROM particles, as the normalized root mean square (RMS) of the differences between their spanwise coordinates at the channel outlet, $\xi_{\text{out}}$ (with $\xi_{\text{out}} = 0 $ if the particle deposits on the left wall of the channel), and their spanwise coordinates at the inlet, $\xi_{\text{in}}$. 
%In figure \ref{f7}b-c, we show the difference in RMS deviation between the healthy tract and the tumor tract, as a function of the particle diameters. The change in vortex properties owing to anatomical distortion has the greatest impact on the trajectories of particles with diameters $\unit[5]{\mu m} < d < \unit[20]{\mu m}$. Comparing panel (a) of figure~\ref{f7} with panels (b, c), the ROM-trend, irrespective of whether we employ low- or high-resolution vortex idealizations, agrees well with the full-scale numerical projections.

% Figure 7

\section{Remarks and Conclusions}\label{discussion}

The anatomy of the upper airway generates strong vortices in the glottic and subglottic regions during inhalation.  These vortical structures may significantly influence the deposition of airborne droplets or particles in these regions.  The increased vortex strength and altered flow patterns resulting from addition of nodule lead to enhanced particle deposition in the affected regions.
A reduced-order model of this flow --- here consisting of a 2D potential flow with slip boundaries and a simplified Maxey-Riley model of particle motion, with parameters guided by high-fidelity computations --- is found to effectively capture particle transmission trends as a signature of the throat geometry and the corresponding vortices.  These results show that the primary influence of the vortical structures on particle transport comes from the overall magnitude of vorticity in this region, not the details of structure and orientation.  This observation is central for understanding the behavior of inhaled particles in individuals with airway abnormalities and for designing effective therapeutic strategies.

The ROM presented here can estimate the influence of airway obstructions on airflow and particle transport through simple vortex modulation, providing a valuable tool for comprehensive parametric analysis of intra-airway transport, thus aiding in diagnosing and treating respiratory conditions and in planning aerosolized targeted drug delivery. 
The ROM may also be used to inform data-driven strategies for rapid and precise evaluation in situ. By leveraging real-time imaging and the resulting geometric parameters, reduced-order modeling can enhance decision-making in clinical settings, enabling prompt assessment of how anatomical variations may impact airflow and particle deposition. This capability is particularly beneficial for personalized medicine --- optimizing treatment procedures and improving therapeutic outcomes based on patient-specific anatomy.\\

%Future work will involve comprehensive experiments covering a broad parameter range, focusing on high-resolution 3D flow field characterization and tracking particle trajectories, velocities, and accelerations. This will enable the study of instability mechanisms, turbulence dynamics, boundary layer modulation, particle deposition, and the effects of irregular breathing, as well as particle Stokes and Reynolds numbers. Extending the ROM to irregular cavities will also enhance model fidelity and insights.

%%%%%%%%%%%%%%%%%%%%%%%%%%%%%%%%%%%%%%%%%%%%%%%%%%%%%%%%%%%%%%%%%%%%%%%%%%%%%%%%%%%%%%%%%%%%%%%%%%%%%%%%%%%%%%%%%%%%%%%%%%%%%%%%%%%%%%%%%%%%%%%%%%%%%%%%%%%%%%%%%%%%%%%%%%%%%%%%%%%%%%%

%\vspace{2mm}

\small
\noindent \textbf{Acknowledgements.} SB acknowledges the \href{https://www.nsf.gov/awardsearch/showAward?AWD_ID=2339001&HistoricalAwards=false}{NSF CAREER Grant No.~CBET 2339001} (FD Program) as support for this work. The authors also thank Julia Kimbell at the School of Medicine, UNC Chapel Hill, for granting access to existing, de-identified upper airway scans.

\vspace{2mm}
\noindent \textbf{Data sharing.} Supplemental info (data and codes) are available on-request via \href{https://jackssdstate-my.sharepoint.com/:f:/g/personal/saikat_basu_sdstate_edu/EgxtOhkKTfJAq5Zyo6IhCPYB-hQUDse7ELspczga72WZtA?e=dJQlb0}{OneDrive}.

\vspace{2mm} 
\noindent {\textbf{Declaration of Interests.}} The authors report no conflict of interest.

%\normalsize

%\vspace{-3.65mm}
\bibliographystyle{jfm}
% Note the spaces between the initials
\bibliography{ReferencesJFM2024}

\end{document}